\documentclass[10pt]{article}
\usepackage{latexsym,graphicx}
\newcommand{\be}{\begin{equation}}
\newcommand{\ee}{\end{equation}}
\def\n{\noindent}
\catcode `\@=11
\catcode `\@=12
\begin{document}
\begin{center}
\large{\bf {Bianchi Type-I Cosmological Models with Variable G and $\Lambda$-Term in General
 Relativity}} \\
\vspace{10mm}
\normalsize{J. P. Singh $^1$, Anirudh Pradhan $^2$ and Ajay Kumar singh $^{3}$} \\
\normalsize{$^{1}$ Department of Mathematical Sciences, A. P. S. University, Rewa-486 003, 
(M. P.), India \\
E-mail : jpsinghmp@lycos.com}\\
\vspace{5mm}
\normalsize{$^{2,3}$ Department of Mathematics, Hindu Post-graduate College, 
Zamania-232 331, Ghazipur, India. \\
$^{2}$ E-mail : pradhan@iucaa.ernet.in}\\
%\normalsize{}\\
\end{center}
\vspace{10mm}
%\date{}
%\maketitle
\begin{abstract} 
Einstein's field equations with variable gravitational and cosmological ``constant'' are considered 
in presence of perfect fluid for Bianchi type-I space-time. Consequences of the four cases of the 
phenomenological decay of $\Lambda$ have been discussed which are consistent with observations. The 
physical significance of the cosmological models have also been discussed. 
\end{abstract}
\smallskip
\n PACS: 98.80.Es, 98.80.-k \\
\n Keywords : Cosmology, Variable cosmological term, Perfect fluid models\\
%\newpage
%%%%%%%%%%%%%%%%%%%%%%%%%%%%%%%%%%%%%%%%%%%%%%%%%%%%%%%%%%%%%%%%%%%%%%%%%%%%%%%%%%%
%%%%%%%%%%%%%%%%%%%%%%%%%%%%%%%   SECTION 1  %%%%%%%%%%%%%%%%%%%%%%%%%%%%%%%%%%%%%%
\section{Introduction}
\noindent
There are significant observational evidence that the expansion of the Universe is undergoing 
a late time acceleration (Perlmutter et al. 1997, 1998, 1999; Riess et al. 1998, 2004; 
Efstathiou et al. 2002; Spergel et al. 2003; Allen et al. 2004; Sahni and Starobinsky 2000; 
Peebles and Ratra 2003; Padmanabhan 2003; Lima 2004). This, in other words, amounts to 
saying that in the context of Einstein's general theory of relativity some sort of dark energy, 
constant or that varies only slowly with time and space dominates the current composition of cosmos. 
The origin and nature of such an accelerating field poses a completely open question. The 
main conclusion of these observations is that the expansion of the universe is accelerating. 
\newline
\par
Among many possible alternatives, the simplest and most theoretically appealing possibility for 
dark energy is the energy density stored on the vacuum state of all existing fields in the universe, 
i.e., $\rho_{v} = \frac{\Lambda}{8\pi G}$, where $\Lambda$ is the cosmological constant. However, 
a constant $\Lambda$ cannot explain the huge difference between the cosmological constant inferred 
from observation and the vacuum energy density resulting from quantum field theories. In an attempt to 
solve this problem, variable $\Lambda$ was introduced such that $\Lambda$ was large in the early 
universe and then decayed with evolution (Dolgov 1983). Cosmological scenarios with a time-varying 
$\Lambda$ were proposed by several researchers. A number of models with different decay laws for the 
variation of cosmological term were investigated during last two decades (Chen and Hu 1991; Pavon 1991; 
Carvalho, Lima and Waga 1992; Lima and Maia 1994; Lima and Trodden 1996; Arbab and Abdel-Rahaman 
1994; Vishwakarma 2001, Cunha and Santos 2004; Carneiro and Lima 2005). 
\newline
\par 
On the other hand, numerous modifications of general relativity to allow for a variable $G$ based on 
different arguments have been proposed (Wesson 1980). Variation of $G$ has many interesting consequences 
in astrophysics. Canuto and Narlikar (1980) have shown that $G$-varying cosmology is consistent with 
whatsoever cosmological observations available at present. A modification linking the variation of 
$G$ with that of variable $\Lambda$-term  has been considered within the framework of general 
relativity by a number of workers (Kallingas et al. 1992; Abdel-Rahaman 1990; Berman 1991; 
Beesham 1986). This modification is appealing as it leaves the form of Einstein's equations formally 
unchanged by allowing a variation of $G$ to be accompanied by a change in $\Lambda$. Cosmological 
models with time-dependent $G$ and $\Lambda$ in the solutions $\Lambda \sim
R^{-2}, \Lambda \sim t^{-2}$, were first obtained by Bertolami (1986). The cosmological models with 
variable $G$ and $\Lambda$ have been recently studied by several authors
(Arbab 2003; Sistero 1991; Sattar and Vishwakarma 1997; Pradhan et al.,
2001, 2002, 2005, 2007; Singh et al., 2006, 2007).    
\newline
\par 
Another important quantity which is supposed to be damped out in the course of cosmic evolution 
is the anisotropy of the cosmic expansion. Theoretical arguments and recent experimental data 
support the existence of an anisotropic phase that approaches an isotropic one. Therefore, it 
makes sense to consider the models of the universe with anisotropic background in the presence 
of dark energy. 
\newline
\par 
The simplest of anisotropic models are Bianchi type-I homogeneous models whose spatial sections 
are flat but the expansion or contraction rate are direction dependent. For studying the possible 
effects of anisotropy in the early universe on present day observations many researchers (Huang 1990; 
Chimento et al. 1997; Lima 1996; Lima and Carvalho 1994; Pradhan et al. 2004, 2006; Saha 2005, 2006) have 
investigated Bianchi type-I models from different point of view.
\newline
\par 
In the present article, we present a new class of solutions to Einstein's field equations with 
variable $G$ and $\Lambda$ in Bianchi type-I space-time in the presence of a perfect fluid. Consequences 
of the following four cases of the phenomenological decay of $\Lambda$ have been discussed:
$$ Case ~ 1: \Lambda \sim H^{2}, $$
$$ Case ~ 2: \Lambda \sim H, $$
$$ Case ~ 3: \Lambda \sim \rho, $$
$$ Case ~ 4: \Lambda \sim R^{-2}, $$
where $H$, $\rho$, $R$ are respectively the Hubble parameter, energy density and average scale factor 
of the Bianchi type-I metric. The dynamical laws proposed for the decay of $\Lambda$ have been widely 
studied by Chen and Wu (1990), Carvalho et al. (1992), Schutzhold (2002), Vishwakarma (2000), Arbab 
(1997, 1998) to name only a few.
%%%%%%%%%%%%%%%%%%%%%%%%%%%%%%%%%%%%%%%%%%%%%%%%%%%%%%%%%%%%%%%%%%%%%%%%%%%%%%%%%%%
%%%%%%%%%%%%%%%%%%%%%%%%%%%%%%%  SECTION 2  %%%%%%%%%%%%%%%%%%%%%%%%%%%%%%%%%%%%%%%%
\section{The Metric, Field  Equations and Solutions}
We consider the space-time admitting Bianchi type-I group of motion in the form 
\begin{equation} 
\label{eq1}  
ds^{2} = - dt^{2} + A^{2}(t) dx^{2} + B^{2}(t)dy^{2} + C^{2}(t)dz^{2}.
\end{equation} 
We assume that the cosmic matter is represented by the energy-momentum tensor of a perfect fluid
\begin{equation} 
\label{eq2}
T_{ij} = (\rho + p) v_{i} v_{j} + p g_{ij},
\end{equation} 
where $\rho$, $p$ are the energy density, thermodynamical pressure and $v_{i}$ is the four-velocity 
vector of the fluid satisfying the relation
\begin{equation} 
\label{eq3}
v_{i}v^{i} = - 1.
\end{equation} 
The Einstein's field equations with time-dependent $G$ and $\Lambda$ are 
\begin{equation} 
\label{eq4}
R_{ij} - \frac{1}{2}Rg_{ij} = - 8\pi G(t) T_{ij} + \Lambda(t) g_{ij}. 
\end{equation} 
For the metric (\ref{eq1}) and energy-momentum tensor (\ref{eq2}) in comoving system of coordinates, 
the field equation (\ref{eq4}) yields
\begin{equation} 
\label{eq5}
8\pi G p - \Lambda = -\frac{\ddot{B}}{B} - \frac{\ddot{C}}{C} - \frac{\dot{B} \dot{C}}{BC},
\end{equation} 
\begin{equation} 
\label{eq6}
8\pi G p - \Lambda = -\frac{\ddot{A}}{A} - \frac{\ddot{C}}{C} - \frac{\dot{A} \dot{C}}{AC},
\end{equation} 
\begin{equation} 
\label{eq7}
8\pi G p - \Lambda = -\frac{\ddot{A}}{A} - \frac{\ddot{B}}{B} - \frac{\dot{A} \dot{B}}{AB},
\end{equation} 
\begin{equation} 
\label{eq8}
8\pi G \rho + \Lambda = \frac{\dot{A} \dot{B}}{AB} + \frac{\dot{B} \dot{C}}{BC} + 
\frac{\dot{A} \dot{C}}{AC}.
\end{equation} 
In view of vanishing divergence of Einstein tensor, we get
\begin{equation} 
\label{eq9}
8\pi G\Big[\dot{\rho} + (\rho + p)\left(\frac{\dot{A}}{A} + \frac{\dot{B}}{B} + 
\frac{\dot{C}}{C}\right)\Big] + 8\pi \rho \dot{G} + \dot{\Lambda} = 0.
\end{equation} 
The usual energy conservation equation $T^{j}_{i;j} = 0$ yields
\begin{equation} 
\label{eq10}
\dot{\rho} + (\rho + p)\left(\frac{\dot{A}}{A} + \frac{\dot{B}}{B} + \frac{\dot{C}}{C}\right) = 0.
\end{equation} 
Equation (\ref{eq9}) together with (\ref{eq10}) puts $G$ and $\Lambda$ in some sort of coupled 
field given by
\begin{equation} 
\label{eq11}
8\pi \rho \dot{G} + \dot{\Lambda} = 0.
\end{equation}
Here and elsewhere a dot stands for ordinary time-derivative of the concerned quantity. From equation 
(\ref{eq11}) one concludes that when $\Lambda$ is constant or $\Lambda = 0$, $G$ turns out to be 
constant. \\
Let $R$ be the average scale factor of Bianchi type-I universe i.e. 
\begin{equation} 
\label{eq12}
R^{3} = \sqrt{-g} = ABC.
\end{equation} 
From equations (\ref{eq5}), (\ref{eq6}) and (\ref{eq7}), we obtain
\begin{equation} 
\label{eq13}
\frac{\dot{A}}{A} - \frac{\dot{B}}{B} = \frac{k_{1}}{R^{3}},
\end{equation} 
and
\begin{equation} 
\label{eq14}
\frac{\dot{B}}{B} - \frac{\dot{C}}{C} = \frac{k_{2}}{R^{3}}.
\end{equation}
On integration equations (\ref{eq13}) and (\ref{eq14}) give
$$ A = m_{1}~ R~ ~ \exp \left[\frac{(2k_{1} + k_{2})}{3}\int{\frac{dt}{R^{3}}}\right],$$
$$ B = m_{2}~ R~ ~ \exp \left[\frac{(k_{2} - k_{1})}{3}\int{\frac{dt}{R^{3}}}\right],$$
\begin{equation} 
\label{eq15}
C = m_{3}~ R ~ ~ \exp \left[-\frac{(k_{1} + 2k_{2})}{3}\int{\frac{dt}{R^{3}}}\right],
\end{equation}
where $k_{1}$, $k_{2}$, $m_{1}$, $m_{2}$, $m_{3}$ are arbitrary constants of integration 
satisfying 
$$m_{1}m_{2}m_{3} = 1.$$
Similar expressions as (\ref{eq15}) have also been established by Saha (2005).\\

Hubble parameter $H$, volume expansion $\theta$, shear $\sigma$ and deceleration parameter $q$ 
are given by
$$\theta = 3H = 3\frac{\dot{R}}{R},$$
$$\sigma = \frac{k}{\sqrt{3}R^{3}}, ~ ~ k>0, \mbox {(constant)} $$
$$q = -1 - \frac{\dot{H}}{H^{2}}.$$
Equations (\ref{eq5})-(\ref{eq8}) and (\ref{eq10}) can be written in terms of $H$, $\sigma$ and $q$ as 
\begin{equation} 
\label{eq16}
8\pi G p = H^{2}(2q - 1) - \sigma^{2} + \Lambda,
\end{equation}
\begin{equation} 
\label{eq17}
8\pi G \rho = 3H^{2} - \sigma^{2} - \Lambda,
\end{equation}
\begin{equation} 
\label{eq18}
\dot{\rho} + 3(\rho + p)\frac{\dot{R}}{R} = 0.
\end{equation}
It is to note that energy density of the universe is a positive quantity. It is believed that at the 
early stages of the evolution when the average scale factor $R$ was close to zero, the energy density 
of the universe was infinitely large. On the other hand, with the expansion of the universe i.e. with 
increase of $R$, the energy density decreases and an infinitely large $R$ corresponds to a $\rho$ 
close to zero. In that case from  (\ref{eq17}), we obtain
\begin{equation} 
\label{eq19}
3H^{2} - \Lambda \to 0.
\end{equation}
From equation (\ref{eq19}) one concludes that: \\

(i) $\Lambda$ is essentially non-negative, \\

(ii) in absence of a $\Lambda$-term beginning from some value of $R$, the evolution of the 
universe becomes stand-still i.e. $R$ becomes constant since $H$ becomes zero, \\

(iii) in case of a positive $\Lambda$, the process of evolution of the universe never comes to 
a halt. Moreover, it is believed that the presence of dark energy (given by positive $\Lambda$) 
results in the accelerated expansion of the universe. As far as negative $\Lambda$ is concerned, 
its presence imposes some restriction on $\rho$ i.e $\rho$ can never be small enough to be ignored. 
It means, in that case there exists some upper limit for $R$ as well. It is worth mention here that 
Saha (2006) has also given such conclusion in his paper but his approach was quite different. \\
From equation (\ref{eq17}), we obtain
$$ \frac{\sigma^{2}}{\theta^{2}} = \frac{1}{3} - \frac{8\pi G \rho}{\theta^{2}} - \frac{\Lambda}
{\theta^{2}}.$$  
Therefore, $0 \leq  \frac{\sigma^{2}}{\theta^{2}} \leq \frac{1}{3}$ and $0 \leq 
\frac{8\pi G \rho}{\theta^{2}} \leq \frac{1}{3}$ for $\Lambda \geq 0$. 

Thus, the presence of a positive $\Lambda$ puts restriction on the upper limit of anisotropy whereas 
a negative $\Lambda$ contributes to the anisotropy. 

From equation (\ref{eq16}), we obtain
$$ \frac{d\theta}{dt} = -\frac{3}{2}\{8\pi G p + 3H^{2} - \Lambda + \sigma^{2}\} $$
Thus for negative $\Lambda$, the universe will always be in decelerating phase whereas a positive $
\Lambda$ will slow down the rate of decrease. Also $\sigma_{4} = -3\sigma H$ implying that $\sigma$ 
decreases in an evolving universe and for infinitely large value of $R$, $\sigma$ becomes negligible.

Equations (\ref{eq5}) - (\ref{eq8}) and (\ref{eq11}) together with one of the decay laws for $\Lambda$
given by cases (1) - (4) supply six equations in seven unknown functions of time $A$, $B$, 
$C$, $\rho$, $p$, $\Lambda$ and $G$. To have deterministic solutions, we require one more condition. 
For this purpose, we assume that the volume expansion $\theta$ is proportional to eigen values of shear 
tensor $\sigma_{ij}$. It is believed that evolution of one parameter should also be responsible for the 
evolution of the others (Vishwakarma, 2005). Following Roy and Singh (1985), we take the volume expansion 
$\theta$ having a constant ratio to the anisotropy in the direction of unit space-like vector 
$\lambda^{i}$ i.e. $\frac{\theta}{\sigma_{ij}\lambda^{i}\lambda^{j}}$ is constant. In general, the above 
condition gives rise to
\begin{equation} 
\label{eq20}
A = B^{m}C^{n},
\end{equation}
where $m$ and $n$ are constants. Using condition (\ref{eq20}) in equations (\ref{eq13}) and 
(\ref{eq14}), we obtain
\[
C = b_{1}(k_{3}t + k_{4})^{\frac{k_{1} - (m - 1)k_{2}}{(m + n + 2)k_{1} - (m - 2n -1)k_{2}}}
~ ~ ~ for ~ ~ \frac{k_1}{k_{2}} \ne \frac{m - 2n - 1}{m + n + 2},
\]
\begin{equation} 
\label{eq21}
= k_{5} ~ \exp{\left[\frac{-k_{2}(m + 1)t}{m + n + 2}\right]} ~ ~ ~ for ~ ~ \frac{k_1}{k_{2}} = 
\frac{m - 2n - 1}{m + n + 2},
\end{equation}
\[
B = b_{2}(k_{3}t + k_{4})^{\frac{k_{1} + k_{2}n}{(m + n + 2)k_{1} - (m - 2n -1)k_{2}}}
~ ~ ~ for ~ ~ \frac{k_1}{k_{2}} \ne \frac{m - 2n - 1}{m + n + 2},
\]
\begin{equation} 
\label{eq22}
= b_{3} ~ \exp{\left[\frac{k_{2}(n + 1)t}{m + n + 2}\right]} ~ ~ ~ for ~ ~ \frac{k_1}{k_{2}} = 
\frac{m - 2n - 1}{m + n + 2},
\end{equation}
provided $m + n \ne 1$. In the above $k_{3}$, $k_{4}$, $k_{5}$ and $b_{1}$, $b_{2}$, $b_{3}$ are 
constants of integration. For these solutions, metric (\ref{eq1}) takes the following forms after 
suitable transformations:
\[
ds^{2} = -dT^{2} + T^{\frac{2(m + n)k_{1} + 2n k_{2}}{(m + n + 2)k_{1} - (m - 2n - 1)k_{2}}} dX^{2}
+ T^{\frac{2k_{1} + 2nk_{2}}{(m + n + 2)k_{1} - (m - 2n - 1)k_{2}}} dY^{2}
\]
\begin{equation} 
\label{eq23}
+ 
T^{\frac{2k_{1} - 2(m - 1) k_{2}}{(m + n + 2)k_{1} - (m - 2n - 1)k_{2}}} dZ^{2} ~ ~ ~ 
for ~ ~ \frac{k_{1}}{k_{2}} \ne \frac{m - 2n - 1}{m + n + 2},
\end{equation}
and
\[
ds^{2} = -dT^{2} + \exp{\left[\frac{2k_{2}(m - n)T}{m + n + 2}\right]}dX^{2} + 
\exp{\left[\frac{2k_{2}(n + 1)T}{m + n + 2}\right]}dY^{2} 
+
\]
\begin{equation} 
\label{eq24}
\exp{\left[-\frac{2k_{2}(m + 1)T}{m + n + 2}\right]}dZ^{2} ~ ~ ~ 
for ~ ~ \frac{k_1}{k_{2}} = \frac{m - 2n - 1}{m + n + 2}.
\end{equation}
%%%%%%%%%%%%%%%%%%%%%%%%%%%%%%%%%%%%%%%%%%%%%%%%%%%%%%%%%%%%%%%%%%%%%%%%%%%%%%%%%%%
%%%%%%%%%%%%%%%%%%%%%%%%%%%%%%%  SECTION 3  %%%%%%%%%%%%%%%%%%%%%%%%%%%%%%%%%%%%%%%%
\section{Discussion}
We now describe the models resulting from different dynamical laws for the decay of $\Lambda$. \\

For the model (\ref{eq23}), average scale factor $R$ is given by
$$ R = T^{\frac{1}{3}}. $$
Volume expansion $\theta$, Hubble parameter $H$ and shear $\sigma$ for 
the model are:
$$ \theta = 3H = \frac{1}{T}, ~ ~ ~  \sigma^{2} = \frac{k^{2}}{3T^{2}}. $$
Thus we see that $\frac{\sigma}{\theta} = \frac{k}{\sqrt{3}}$. Therefore, the model does not approach 
isotropy. If $k$ is small, the models are quasi-isotropic i.e. $\frac{\sigma}{\theta} = 0$.
%%%%%%%%%%%%%%%%%%%%%%%%%%%%%%%%%%%%%%%%%%%%%%%%%%%%%%%%%%%%%%%%%%%%%%%%%%%%%%%%%%%
%%%%%%%%%%%%%%%%%%%%%%%%%%%%%%%  SUBSECTION 3.1  %%%%%%%%%%%%%%%%%%%%%%%%%%%%%%%%%%%%%%%%
\subsection{Case 1 :} 
We consider 
$$\Lambda = 3\beta H^{2},$$
where $\beta$ is a constant of the order of unity. Here $\beta$ represents the ratio between vacuum 
and critical densities. From equations (\ref{eq5}), (\ref{eq8}) and (\ref{eq11}), we obtain  
\begin{equation} 
\label{eq25}
8\pi \rho = \frac{(1 - k^{2} - \beta)}{3k_{0}} T^{-\frac{2(1 - k^{2})}{(1 - k^{2} - \beta)}},
\end{equation}
\begin{equation} 
\label{eq26}
8\pi p = \frac{(1 - k^{2} + \beta)}{3k_{0}} T^{-\frac{2(1 - k^{2})}{(1 - k^{2} - \beta)}},
\end{equation}
\begin{equation} 
\label{eq27}
\Lambda = \frac{\beta}{3T^{2}},
\end{equation}
\begin{equation} 
\label{eq28}
G = k_{0}T^{\frac{2\beta}{(1 - k^{2} - \beta)}}, ~ ~ ~ k_{0} > 0 \mbox{(constant)}.
\end{equation}
We observe that the model has singularity at $T = 0$. It starts with 
a big bang from its singular state and continues to expand till $T = \infty$. At $T = 0$, $\rho$, $p$, 
$\Lambda$, $\theta$ and $\sigma$ are all infinite whereas $G = 0$ for $\beta > 0$ and $G = \infty$ for 
$\beta < 0 $. For infinitely large $T$, $\rho$, $p$, $\Lambda$, $\theta$ and $\sigma$ are all zero 
but $G = \infty$ for $\beta > 0$ and $G = 0$ for $\beta < 0$. We also observe that in the absence of 
cosmological term $\Lambda(\beta = 0)$, $\rho = p$ i.e. matter content turns out to be a stuff fluid. 
For $ \beta > 0$, $p > \rho$ and $p < \rho$ when $\beta < 0$. When $\beta = k^{2} - 1$, $p = 0$.  The 
density parameter $\Omega = \frac{\rho}{\rho_{c}} = 1 - k^{2} - \beta$ implying that $\rho_{c} > \rho$ 
and $\rho_{c} < \rho$ for $\beta > - k^{2}$ and  $\beta < - k^{2}$ respectively whereas $\rho_{c} = 
\rho$ when $\beta = - k^{2}$. The ratio between vacuum and matter densities is given by 
$$\frac{\rho_{v}}{\rho} = \frac{\beta}{1 - k^{2} - \beta}.$$
%%%%%%%%%%%%%%%%%%%%%%%%%%%%%%%%%%%%%%%%%%%%%%%%%%%%%%%%%%%%%%%%%%%%%%%%%%%%%%%%%%%
%%%%%%%%%%%%%%%%%%%%%%%%%%%%%%%  SUBSECTION 3.2  %%%%%%%%%%%%%%%%%%%%%%%%%%%%%%%%%%%%%%%%
\subsection{Case 2 :} 
We now consider 
$$\Lambda = a H,$$
where $a$ is a positive constant of order of $m^{3}$ where $m \approx 150 MeV$ is the energy scale of 
chiral phase transition of QCD (Borges and Carneiro, 2005). For this case, equations (\ref{eq5}), 
(\ref{eq8}) and (\ref{eq11}) yield 
\begin{equation} 
\label{eq29}
8\pi \rho = \frac{(1 - k^{2} - aT)^{2}}{3k_{0}T^{2}},
\end{equation}
\begin{equation} 
\label{eq30}
8\pi p = \frac{(1 - k^{2})^{2} - a^{2}T^{2}}{3k_{0}T^{2}},
\end{equation}
\begin{equation} 
\label{eq31}
\Lambda = \frac{a}{3T},
\end{equation}
\begin{equation} 
\label{eq32}
G = \frac{k_{0}}{1 - k^{2} - aT}.
\end{equation}
The model has singularity at $T = 0$. The model starts from a big bang with  $\rho$, $p$, 
$\Lambda$, $\theta$, $\sigma$  all infinite and $G$ finite. Thereafter $\rho$, $p$, 
$\Lambda$, $\theta$ and $\sigma$ decrease and $G$ increases. When $T = \frac{1 - k^{2}}{a}$, 
we obtain $p = 0$, $\rho = 0$, $\Lambda = \frac{a^{2}}{3(1 - k^{2})}$, $\sigma = \frac{ka}
{\sqrt{3}(1 - k^{2})}$ and $G$ is infinite. As $T \to \infty $, $\rho \sim \frac{a^{2}}{24\pi k_{0}}$, 
 $p \sim -\frac{a^{2}}{24\pi k_{0}}$, and $\theta $, $\sigma$, $G$, $\Lambda$ tend to zero. The 
density parameter $\Omega = 1 - k^{2} - aT$ and the ratio between vacuum and critical densities is 
given by $$ \frac{\rho_{v}}{\rho_{c}} = aT. $$
%%%%%%%%%%%%%%%%%%%%%%%%%%%%%%%%%%%%%%%%%%%%%%%%%%%%%%%%%%%%%%%%%%%%%%%%%%%%%%%%%%%
%%%%%%%%%%%%%%%%%%%%%%%%%%%%%%%  SUBSECTION 3.3  %%%%%%%%%%%%%%%%%%%%%%%%%%%%%%%%%%%%%%%%
\subsection{Case 3 :} 
We now consider 
$$\Lambda = \frac{8\pi \alpha G \rho}{3},$$
where $\alpha$ is a constant. In this case from  equations (\ref{eq5}), (\ref{eq8}) and (\ref{eq11}), 
we obtain 
\begin{equation} 
\label{eq33}
8\pi \rho = \frac{(1 - k^{2})}{k_{0}(\alpha + 3)} T^{-\frac{2(\alpha + 3)}{3}}, ~ ~ ~ \alpha \ne -3,
\end{equation}
\begin{equation} 
\label{eq34}
8\pi p = \frac{(1 - k^{2})(2\alpha + 3)}{k_{0}(\alpha + 3)} T^{-\frac{2(\alpha + 3)}{3}},
\end{equation}
\begin{equation} 
\label{eq35}
\Lambda = \frac{\alpha(1 - k^{2})}{3(\alpha + 3)T^{2}},
\end{equation}
\begin{equation} 
\label{eq36}
G = k_{0}T^{\frac{2\alpha}{3}}.
\end{equation}
This model also starts from a big bang at $T = 0$ with $\rho$, $p$, $\Lambda$, $\theta$, $\sigma$ 
all infinite and $G = 0$ (for $\alpha > 0$) and it evolves to $\rho \to 0$, $p \to 0$, $\theta \to 0$, 
$\sigma \to 0$, $\Lambda \to 0$ and $G \to \infty$ as $T \to \infty$. The density parameter $\Omega$ 
for this model is given by 
$$ \Omega = \frac{3(1 - k^{2})}{\alpha + 3},$$
and the ratio between vacuum and critical densities is obtained as
$$ \frac{\rho_{v}}{\rho_{c}} = \frac{\alpha(1 - k^{2})}{\alpha + 3}. $$  
%%%%%%%%%%%%%%%%%%%%%%%%%%%%%%%%%%%%%%%%%%%%%%%%%%%%%%%%%%%%%%%%%%%%%%%%%%%%%%%%%%%
%%%%%%%%%%%%%%%%%%%%%%%%%%%%%%%  SUBSECTION 3.4  %%%%%%%%%%%%%%%%%%%%%%%%%%%%%%%%%%%%%%%%
\subsection{Case 4 :} 
Finally we consider the case 
$$\Lambda = \frac{\gamma}{R^{2}},$$
where $\gamma$ is a parameter to be determined from the observations. 
In this case from  equations (\ref{eq5}) - (\ref{eq8}) and (\ref{eq11}), 
we obtain 
\begin{equation} 
\label{eq37}
8\pi \rho = \frac{[(1 - k^{2})T^{-\frac{4}{3}} - 3\gamma]^{\frac{3}{2}}}{3k_{0}},
\end{equation}
\begin{equation} 
\label{eq38}
8\pi p = \frac{[(1 - k^{2})T^{-\frac{4}{3}} + 3\gamma]^{\frac{3}{2}}}{3k_{0}},
\end{equation}
\begin{equation} 
\label{eq39}
\Lambda = \gamma T^{-\frac{2}{3}},
\end{equation}
\begin{equation} 
\label{eq40}
G = k_{0}[1 - k^{2} - 3\gamma T^{\frac{4}{3}}]^{-\frac{1}{2}}.
\end{equation}
Here we observe that this model also has singularity at $T = 0$. It starts from a big bang singularity 
with $\rho$, $p$, $\theta$, $\Lambda$, $\sigma$ all infinite but $G$ finite. For $\Lambda > 0$ i.e. 
$\gamma > 0$, $\rho$ becomes zero at $T = \left(\frac{1 - k^{2}}{3\gamma}\right)^{\frac{3}{4}}$ whereas 
for $\Lambda < 0$ i.e. $\gamma <0$, $p = 0$ at $T = \left(\frac{k^{2} - 1}{3\gamma}\right)^{\frac{3}{4}}$. 
As $T \to \infty$, $\theta$, $\sigma$, $\Lambda$ and $G$ become zero but $\rho$ and $p$ become finite. 
The density parameter $\Omega$ for this model is given by 
$$ \Omega = 1 - k^{2} - 3 \gamma T^{\frac{4}{3}}.$$
The ratio between vacuum and critical densities is given by 
$$ \frac{\rho_{v}}{\rho_{c}} = 3\gamma T^{\frac{4}{3}}. $$   
The model (\ref{eq24}) is not of much interest since it reduces to a static solution.
%%%%%%%%%%%%%%%%%%%%%%%%%%%%%%%%%%%%%%%%%%%%%%%%%%%%%%%%%%%%%%%%%%%%%%%%%%%%%%%%%%%
%%%%%%%%%%%%%%%%%%%%%%%%%%%%%%%  SECTION 4  %%%%%%%%%%%%%%%%%%%%%%%%%%%%%%%%%%%%%%%%
\section{Conclusion}
In this paper, we have presented a class of solutions to Einstein's field 
equations with variable $G$ and $\Lambda$ in Bianchi type-I space-time in the presence of a 
perfect fluid. In some cases, it is observed that $G$ is an increasing
function of time. When the universe is required to have expanded from a finite
minimum volume, the critical density assumption and conservation of
energy-momentum tensor dictate that $G$ increases in a perpetually expanding 
universe. The possibility of an increasing $G$ has been suggested by several authors.\\

We would like to mention here that Beesham (1994), Lima and Carvalho (1994), Kallingas et al. 
(1995) and Lima (1996) have also derived the Bianchi type I cosmological models with variable 
$G$ and $\Lambda$ assuming a particular form for $G$. These models have some similarities with 
our model (\ref{eq23}) in the cases (1) and (3) only. But our derived results differ from 
these models in the sense that the both of these are constrained by the equation of state 
whereas we have neither assumed equation of state nor particular form of $G$. \\  
 
The behaviour of the universe in our  models will be determined by the cosmological 
term $\Lambda$ ; this term has the same effect as a uniform mass density $\rho_{eff} 
= - \Lambda / 4\pi G$, which is constant in space and time. A positive value of 
$\Lambda$ corresponds to a negative effective mass density (repulsion). Hence, we 
expect that in the universe with a positive value of $\Lambda$, the expansion will 
tend to accelerate; whereas in the universe with negative value of $\Lambda$, 
the expansion will slow down, stop and reverse.  Recent cosmological 
observations (Garnavich et al. 1998; Perlmutter et al. 1997, 1998, 1999; 
Riess et al. 1998, 2004;  Schmidt et al. 1998) suggest the existence of 
a positive cosmological constant $\Lambda$ with the magnitude 
$\Lambda (G\hbar/c^{3})\approx 10^{-123}$. These observations on magnitude and 
red-shift of type Ia supernova suggest that our universe may be an accelerating 
one with induced cosmological density through the cosmological $\Lambda$-term. 
Thus, our models are consistent with the results of recent observations. \\
\section*{Acknowledgements} 
The authors are grateful to the referee for his valuable comments.
  
%\newline
%\newline

\end{document}